\documentclass[aps,pre,reprint,floatfix,superscriptaddress]{revtex4-1}
\usepackage{graphicx}
\usepackage[latin1]{inputenc}
\usepackage{amsmath}
\usepackage{amsfonts}
\usepackage{amssymb}
\usepackage{xcolor}
\usepackage{physics}

\begin{document}

\title{Compression-induced crossovers for the ground-state of classical dipole lattices on a M\"{o}bius strip}

\date{\today}

\author{A.~Siemens}
\email{asiemens@physnet.uni-hamburg.de}
\affiliation{Zentrum f\"ur Optische Quantentechnologien, Fachbereich Physik, Universit\"at Hamburg, Luruper Chaussee 149, 22761 Hamburg Germany} 
\author{F.~A.~O.~Silveira}
\affiliation{Departamento de Física, UNESP - Universidade Estadual Paulista - Av. 24A,1515, Bela Vista, CEP: 13506-900 - Rio Claro - S\~{a}o Paulo - Brazil } 
\author{P.~Schmelcher}
\email{pschmelc@physnet.uni-hamburg.de}
\affiliation{Zentrum f\"ur Optische Quantentechnologien, Fachbereich Physik, Universit\"at Hamburg, Luruper Chaussee 149, 22761 Hamburg Germany}
\affiliation{Hamburg Center for Ultrafast Imaging, Universit\"at Hamburg, Luruper Chaussee 149, 22761 Hamburg Germany}

\begin{abstract}

\noindent We explore the ground state properties of a lattice of classical dipoles spanned on the surface of a M\"{o}bius strip. 
The dipole equilibrium configurations depend significantly on the geometrical parameters of the M\"{o}bius strip, as well as on the lattice dimensions. 
As a result of the variable dipole spacing on the curved surface of the M\"{o}bius strip, the ground state can consist of multiple domains with different dipole orientations which are separated by domain walls. 
We analyze in particular the dependence of the ground state dipole configuration on the width of the M\"{o}bius strip and highlight two crossovers in the ground state that can be correspondingly tuned. 
A first crossover changes the dipole lattice from a phase which resists compression to a phase that favors it. 
The second crossover leads to an exchange of the topological properties of the two involved domains. 
We conclude with a brief summary and an outlook on more complex topologically intricate surfaces. 

\end{abstract}

\maketitle

\section{Introduction}

Long-range dipole interactions are ubiquitous in physics and appear in a wide range of systems, ranging from atomic setups, such as Rydberg arrays \cite{samajdar2020} or dipolar quantum gases \cite{chomaz2023,ruhman2012}, to solid state systems, such as magnets \cite{johnston2016} or ferroelectrics \cite{qiao2021,kanzig1957}. 
Especially in crystalline lattices, such as in ferroelectric (FE) materials, the anisotropic character of the interaction can lead to the formation of complex ordered phases \cite{martin2017}. 
For example, the degeneracy of the ground-state (GS) configuration, i.e. the invariance of the energy under inversion of all dipoles, can lead to the formation of local domains separated by a domain wall (DW) \cite{meier2022}.  
For FE materials, experiments have shown a great control of these DWs, allowing for controlled shifts and even the controlled creation or annihilation of domains \cite{meier2022,sharma2017}.  
Due to this direct control of the dipole configurations, FE materials have been used for applications, such as smart sensors, capacitors, transducers, actuators, energy harvesting devices, and non-volatile memories \cite{meier2022,wang2022,muralt2000,sharma2017}.

The ordered phases emerging in lattice systems of interacting dipoles can significantly depend on the underlying lattice geometry: 
In certain lattice geometries, the ground state becomes continuously degenerate \cite{schildknecht2019,brankov1987} and allows for continuous transformations between different ground state configurations \cite{debell1997,feldmann2008}. 
Other examples include spin glass phases emerging in disordered systems \cite{mydosh1996}, and the suppression of long-range order in lattices exhibiting geometric frustration \cite{ramirez2001,melchy2009,gaulin1994}. 
Besides these well known examples, interesting geometry-dependent effects can also be found in lattice systems exhibiting mixed dimensionality: 
Already in a simple 1D setup consisting of dipoles that are spaced equidistant along a helical path, the ground state can be classified by a complex self-similar bifurcation diagram that depends on the helix geometry \cite{siemens2022}. 
For dipoles that are arranged on 2D surfaces, the curved geometries can enforce the presence of topological defects, as can be seen for self-assembling dipoles on a sphere \cite{lieu2020}. 
Furthermore, it has been demonstrated that dipole lattices on a 2D curved surface can exhibit domains and domain walls in their ground state \cite{siemens2023}. 
Here, we build upon these results and further investigate the properties of classical dipole lattices in curved geometries. 
Specifically, we are interested in the effects arising when a dipole lattice is spanned on a curved surface that is topologically non-trivial. 

It has been demonstrated that spatial curvature or mixed dimensionality by itself can lead to a variety of intriguing (and often counter-intuitive) effects. 
Already for (isotropic) Coulomb-interacting particles confined to a curved 1D path a plethora of highly non-trivial static \cite{schmelcher2011,plettenberg2017,zampetaki2018,siemens2020} and dynamic \cite{zampetaki2013,zampetaki2015,zampetaki2015a,zampetaki2017,siemens2021} effects can emerge. 
Furthermore, in geometries that are topologically non-trivial, such as the M\"{o}bius strip, the surface topology can induce effects that are absent in corresponding topologically trivial systems \cite{beugeling2014,flouris2022}. 
This motivates us to investigate the ground state properties of a lattice of classical dipoles spanned on the surface of a M\"{o}bius strip. 
We find that a compression of the strip can lead to two distinct crossovers in the ground state of the embedded dipole lattice that are detected as peaks in the compression module. 
One of the crossovers is connected to the curvature-dependent changes of the dipole configurations and corresponds to a change of the system from resisting to favoring compression. 
The second crossover has its origin in the non-trivial M\"{o}bius strip topology, and corresponds to a change of the topological properties of the ground state domains.

Our work is structured as follows: 
The description of our setup is provided in Sec. \ref{Sec:2_math}. 
An overview of the stable GS equilibrium configurations, as well as their dependence on the system parameters, is given in Sec. \ref{Sec:3_eq_conf}. 
In Sec. \ref{Sec:4_crossover}, the two crossovers are discussed. 
Finally, we present in \ref{Sec:5_end} our brief summary and conclusions as well as an outlook.

\section{Lattice of dipoles on a M\"{o}bius strip}\label{Sec:2_math}

\begin{figure}
	\includegraphics[width=\columnwidth]{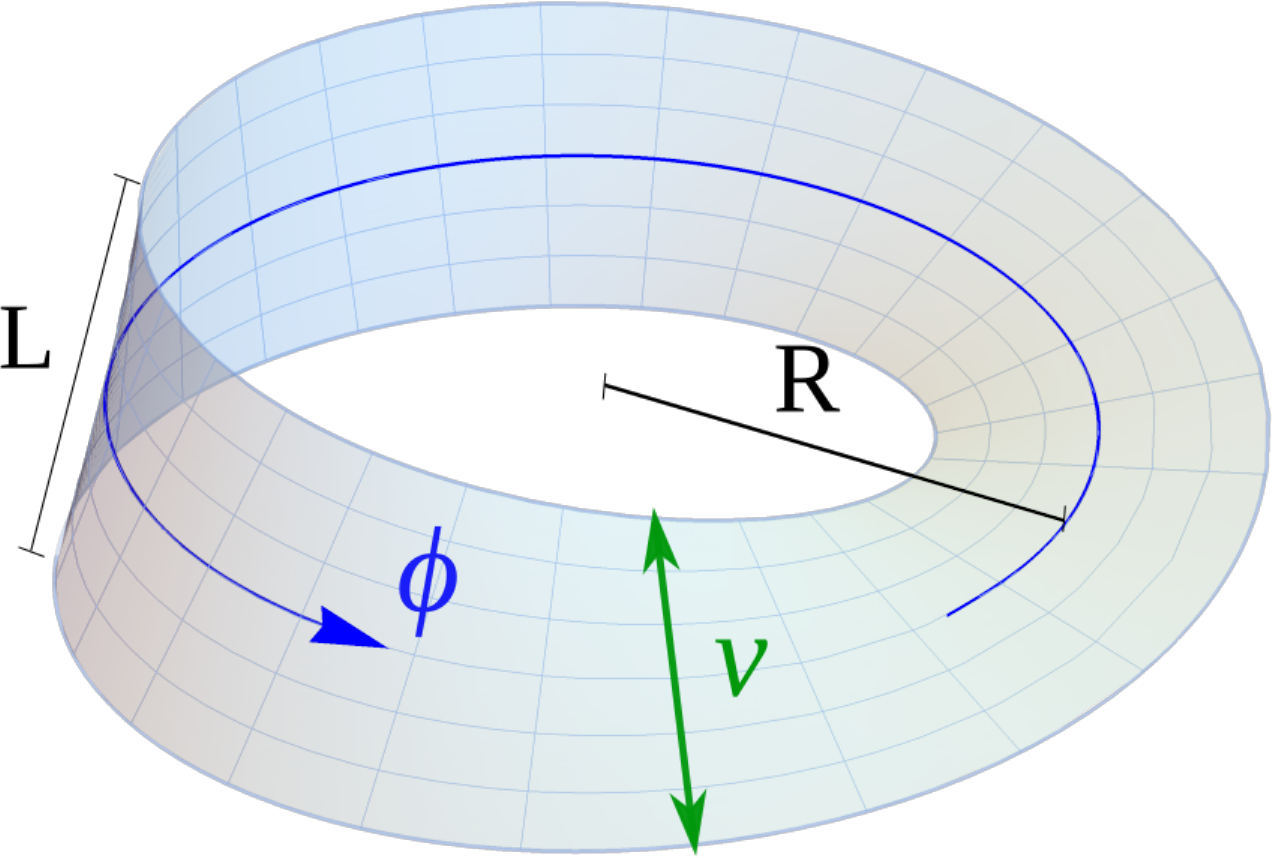}
	\caption{\label{Fig:1}Visualization of the M\"{o}bius strip surface and the parameters of Eq. (\ref{Eq:Parametization}). The mesh grid corresponds to a lattice with $6 \cross 26$ lattice points. }
\end{figure}

We consider a lattice of classical dipoles spanned on the surface a M\"{o}bius strip. 
Each point on the  M\"{o}bius strip's surface can be expressed by a parametric function $\boldsymbol f(u,v)$ given by 
\begin{equation}
	\label{Eq:Parametization}
	\boldsymbol f(\phi,v):=\left(
	\begin{array}{c}
		\left[ R+ v\cos(\phi/2) \right]\cos(\phi) 
		\\
		\left[ R+ v\cos(\phi/2) \right]\sin(\phi)
		\\
		v\sin\left(\phi/2\right)
	\end{array}
	\right),
\end{equation}
where $u$ and $v$ are the parametric (i.e. the internal) coordinates of the surface, and $R$ is the `radius' of the center circle of the M\"{o}bius strip. 
For $\phi\in[0,2\pi)$ and $v\in[-\frac{L}{2},\frac{L}{2}]$, Eq. (\ref{Eq:Parametization}) produces a M\"{o}bius strip with a width $L$. 
Before describing the dipole lattice on the M\"{o}bius strip surface, it is helpful to introduce the unit vectors
\begin{equation}
	\begin{aligned}
		\textbf{e}_{\phi} =& \dfrac{\partial \boldsymbol f(\phi,v)}{\partial \phi} / \left|\left|\dfrac{\partial \boldsymbol f(\phi,v)}{\partial \phi}\right|\right|
		\\
		\textbf{e}_v =& \dfrac{\partial \boldsymbol f(\phi,v)}{\partial v} / \left|\left|\dfrac{\partial \boldsymbol f(\phi,v)}{\partial v}\right|\right| .
	\end{aligned}
\end{equation}
At every point $f(\phi,v)$ the two unit vectors $\textbf{e}_\phi$ and $\textbf{e}_v$ are orthogonal to each other and tangential to the M\"{o}bius strip surface. 
We will from now on respectively refer to $\textbf{e}_\phi$ and $\textbf{e}_v$ as the angular and the radial direction on the M\"{o}bius strip. 
The parametric surface $\boldsymbol f(\phi,v)$ is shown in Fig. \ref{Fig:1}, together with visualizations of the above described parameters.

We now place a grid of $\left(N\times M\right)$ dipoles on the M\"{o}bius strip. 
The grid points are equidistant in the parametric coordinates, with (parametric) lattice constants of $\Delta \phi = 2\pi/N$ and $\Delta v = L/M$. 
Consequently the positions of the lattice points in Euclidean space are given by $\textbf{r}_{nm} = \boldsymbol f\left(n\Delta \phi,m\Delta v-L/2\right)$, where $n\in[1,N]$ and $m\in[1,M]$. 
An example of such a grid with $N=26$ and $M=6$ is visualized on the M\"{o}bius surface in Fig. \ref{Fig:1} (thin gray lines). 
At each position $\textbf{r}_{nm}$, we place a dipole with dipole moment $\textbf{d}_{nm}$. 
The dipoles can freely rotate and interact via dipole-dipole interactions. 
The potential energy $V_{nm}^{ij}$ resulting from the interaction between two dipoles positioned at $\textbf{r}_{nm}$ and $\textbf{r}_{ij}$ is then given by 
\begin{equation}
	V_{nm}^{ij} = \frac{\textbf{d}_{nm}\cdot\textbf{d}_{ij}}{4\pi\epsilon_0 \left(r_{nm}^{ij}\right)^3}-\frac{3\left(\textbf{d}_{nm}\cdot\textbf{r}_{nm}^{ij}\right)\left(\textbf{d}_{ij}\cdot\textbf{r}_{nm}^{ij}\right)}{4\pi\epsilon_0 \left(r_{nm}^{ij}\right)^5} ,
\end{equation}
where $\textbf{r}_{nm}^{ij} = \textbf{r}_{nm} - \textbf{r}_{ij}$ is the (Euclidean) distance vector between the two dipoles, and $r_{nm}^{ij} = |\textbf{r}_{nm}^{ij}|$ the corresponding magnitude. 
The total energy of the system can then be determined by summing up all pairwise interactions $V_{tot} = \sum_{n,m\neq i,j} V_{nm}^{ij}$. 
We are interested in finding the ground state dipole configuration of the lattice, i.e. the configuration that minimizes $V_{tot}$. 
Since the magnitude of the dipole moments $\textbf{d}$ only scales the total energy and does not affect the ground state dipole configuration, we can - without loss of generality - set $d = |\textbf{d}| = 1$. 
This optimization problem then depends on $2MN+4$ parameters, namely the dipole moments $\textbf{d}$, the M\"{o}bius parameters $R$ and $L$, and the (parametric) lattice constants $\Delta u$ and $\Delta v$. 
For the calculation of the GS configurations we consider all-to-all interactions. 
Nevertheless, some of the presented results were obtained using a nearest neighbor (NN) approximation. 
This NN approximation provides a very good approximation of the actual equilibrium configurations for systems where the NN distance is small compared to the curvature radius of the surface \cite{siemens2023}. 
Results based on NN calculations are specifically referred to as such in the text. 
Furthermore, to find the GS configurations we use a principal axis method. 
The principal axis method is a numerical optimization method that does not rely on gradients. 
Instead, the optimizer performs line searches along a set of continuously updated search directions.

\section{GS equilibrium configurations}\label{Sec:3_eq_conf}

\begin{figure}
	\includegraphics[width=\columnwidth]{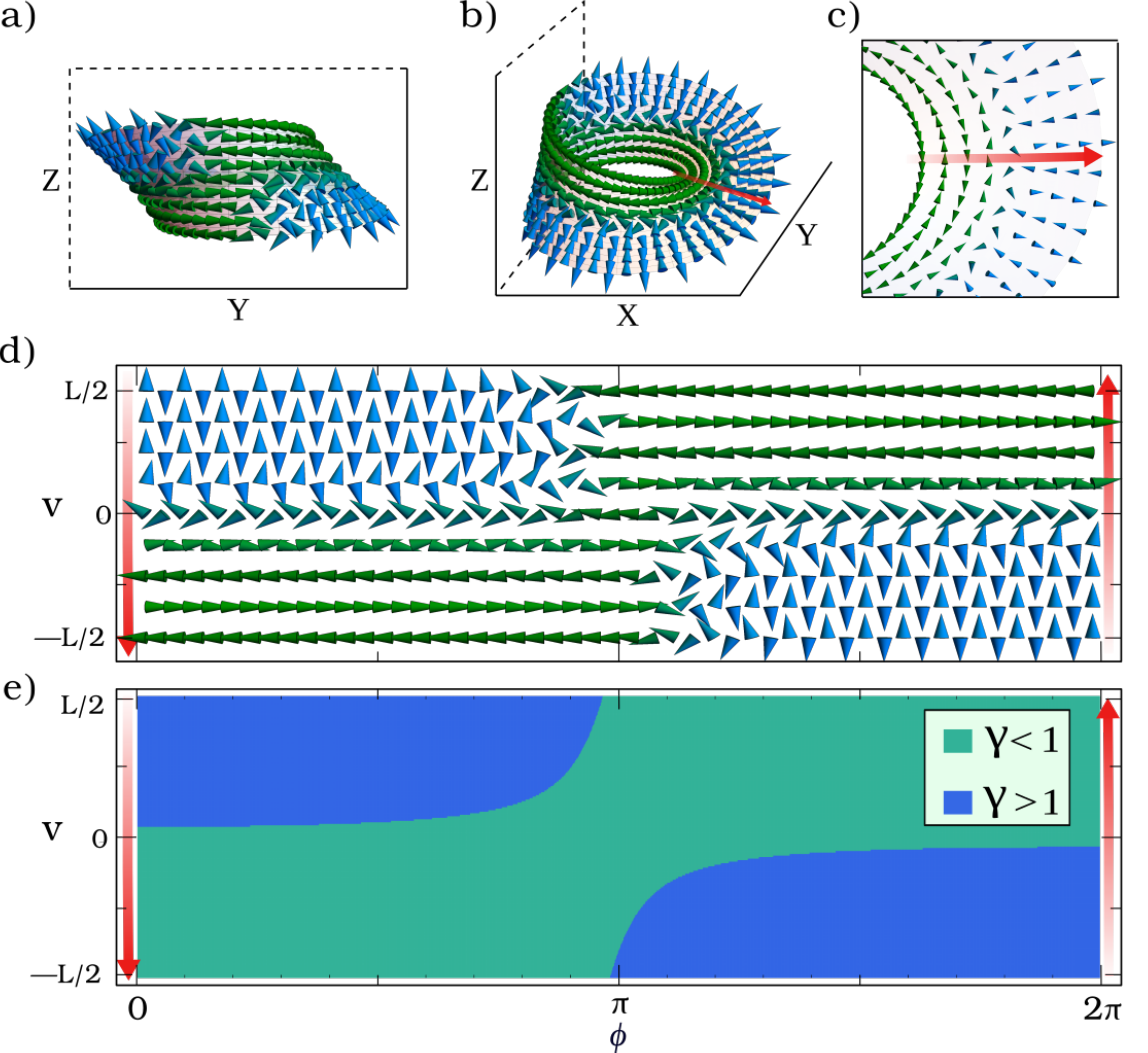}
	\caption{\label{Fig:2} 
	Example ground state dipole configuration for $N=51$, $M=9$, $R=1$, and $L=1.02$. 
	(a-c) Visualizations of the ground state from different viewpoints. 
	(d) Visualization of the ground state dipole alignments within the surface. Dipole positions and alignments are given w.r.t. the parametric coordinates $\phi$ and $v$. 
	(e) A visualization of the parameter $\gamma=a/b$ as a function of the parametric coordinates. See text for details. 
	 }
\end{figure}

The ground state dipole configuration of the above described dipole lattice on a M\"{o}bius strip differs from the well-known ground states of dipole lattices in `flat' geometries. 
This is because the distances between neighboring dipoles on the M\"{o}bius strip depends on the local geometry of $f(\phi,v)$. 
This can be easily seen by calculating the Euclidean distances $a$ and $b$ between neighboring dipoles along the $\textbf{e}_\phi$ and $\textbf{e}_v$ directions
\begin{equation}
	\begin{aligned}
		a(\phi,v)&=\parallel \boldsymbol f(\phi,v+\Delta v) - \boldsymbol f(\phi,v) \parallel,\\
		b(\phi,v)&=\parallel \boldsymbol f(\phi+\Delta \phi, v) - \boldsymbol f(\phi,v) \parallel.
	\end{aligned}
\end{equation}
Inserting the M\"{o}bius parameterization of $\boldsymbol f(\phi,v)$ from Eq. (\ref{Eq:Parametization}) into the above equations yields
\begin{equation}
	a = \Delta v = L/M
\end{equation}
for the Euclidean distance between next neighbors along the radial direction $\textbf{e}_v$, and 
\begin{equation}
\begin{aligned}
	&b^2(\phi,v) = 8 R^2 \cos\left(\frac{\Delta \phi}{4}\right)^2 \\
	&+ v^2 \left[3 + \cos(\phi) + 2 \cos\left(\phi + \frac{\Delta \phi}{2}\right) + 2 \cos(\Delta \phi/2) \right]\\
	&+ v^2 \left[ \cos(\Delta \phi) + \cos(\phi + \Delta \phi)\right] \\ 
	&+ 16 R v \cos\left(\frac{\Delta \phi}{4}\right)^3 \cos\left(\frac{2 \phi + \Delta \phi}{4}\right)
\end{aligned}
\end{equation}
for the Euclidean distance between next neighbors along the angular direction $\textbf{e}_\phi$. 
The impact of such a varying nearest neighbor distance on the dipole equilibrium configurations has been previously studied \cite{siemens2023}. 
Following the nomenclature set in Ref. \cite{siemens2023}, we introduce the parameter $\gamma=a/b$. 
From Ref. \cite{siemens2023}, we know that the dipoles will favor aligning along $\textbf{e}_v$ wherever $\gamma<1$ (i.e. $a<b$), and along $\textbf{e}_\phi$ wherever $\gamma>1$ (i.e. $a>b$). 
If the parameters are chosen such, that in some part of the M\"{o}bius strip $\gamma<1$ and in another part we have $\gamma>1$, the ground state will feature two domains with different dipole orientations separated by a domain wall. 
A detailed description on the properties and the mechanism behind the formation of the domain wall can be found in Ref. \cite{siemens2023}.

An example ground state dipole configuration on the M\"{o}bius strip with $N=51$, $M=9$, $R=1$, and $L=1.02$ is shown in Fig. \ref{Fig:2}(a-c). 
In the figure, the dipoles are colored depending on their orientation: 
Dipoles with $\textbf{d}\parallel\textbf{e}_\phi$ are colored green, whereas all dipoles with $\textbf{d}\parallel\textbf{e}_v$ are colored blue. 
Dipoles with significant alignment normal to the surface, i.e., dipoles for which $\textbf{d}\parallel(\textbf{e}_\phi\cross\textbf{e}_v)$, could not be observed in any of our simulations. 
From now on, we will use the terms angular domain and radial domain to respectively refer to the domains where dipoles are dominantly aligned along $\textbf{e}_\phi$ and $\textbf{e}_v$. 
For better visualization of this ground state, the orientation of the dipoles w.r.t. the parametric coordinates $\phi$ and $v$ is shown in Fig. \ref{Fig:2}(d). 
A corresponding diagram of how the parameter $\gamma(\phi,v)$ changes with the parametric coordinates is shown in Fig. \ref{Fig:2}(e).

From the comparison in Fig. \ref{Fig:2}, as well as from Ref. \cite{siemens2023}, we know that the ground state dipole configuration can be accurately predicted from the parameter $\gamma$, especially when the dipole spacing is small compared to the curvature radius of the surface. 
Therefore, we can get an intuition of the impact of parameter variations on the ground state by analyzing the impact of these changes on $\gamma$. 
For a given dipole lattice of dimension $(N\times M)$, the local value of $\gamma$ can be impacted by the M\"{o}bius parameters $L$ and $R$. 
To get a first impression of the overall behavior of $\gamma$, we expand $b(\phi,v)$ to the first order in $\Delta\phi$ around $\Delta\phi = 0$. 
Since $\Delta\phi = 2\pi/N$, this is a good approximation in the limit of large $N$. 
With this, the parameter $\gamma$ can be approximated as
\begin{equation}\label{Eq:gamma_approx}
	\gamma\simeq
	\frac{L}{\pi}\frac{(N/M)}{\sqrt{4R^2+3v^2+8Rv\cos\left(\frac{\phi}{2}\right)+2v^2\cos(\phi)}}. 
\end{equation}
Consequently, in grids with many lattice points, the parameter $\gamma$ will scale globally when the ratio $N/M$ changes. 
Note that any change of $L$ will also affect the range of the parameter $v\in[-L/2,L/2]$. 
Therefore, it is possible to rewrite Eq. (\ref{Eq:gamma_approx}) such that it depends entirely on the ratio $L/R$ by introducing $v'=2v/L\in[-1,1]$. 
Changing the ratio $L/R$ will not scale $\gamma$ globally. 
Instead, an increase in $L/R$ can either increase $\gamma$ everywhere, or (below a certain value of $L/R$) it can lead to an increase of $\gamma$ in some parts of the lattice and to a decrease of $\gamma$ in other parts.

\begin{figure}
	\includegraphics[width=\columnwidth]{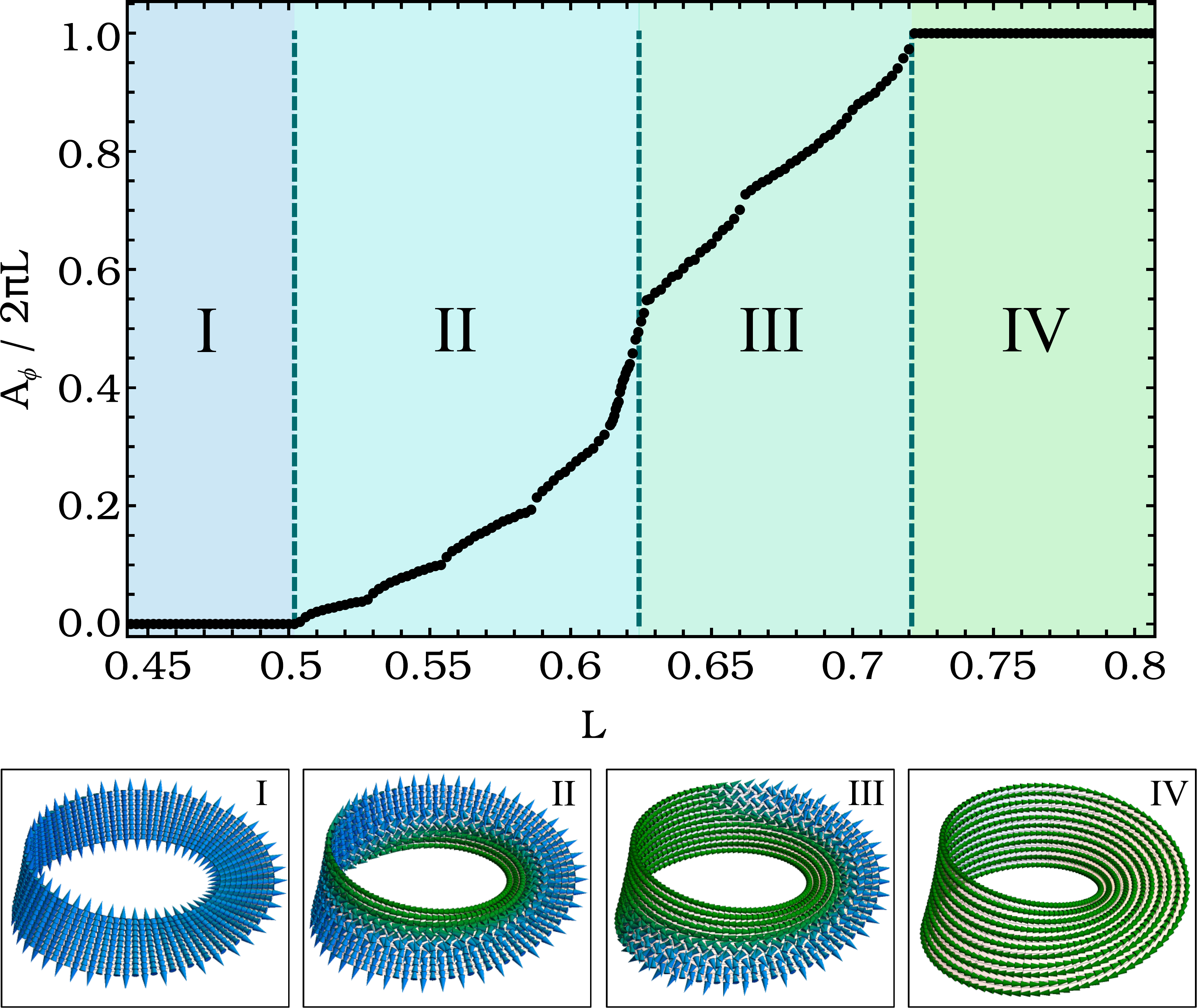}
	\caption{\label{Fig:3} 
		Size $A_\phi$ of the angular domain as a function of the M\"{o}bius strip width $L$ for $N=101$, $M=11$, and $R=1$. 
		All data points were obtained using a nearest neighbor approximation. 
		The lower panel shows example configurations from the four regimes for $L\in[0.47,0.6,0.65,0.83]$ obtained from all-to-all calculations.
		}
\end{figure}

In the following, we will study the ground GS for fixed values of $N/M$ and $R$. 
This reduces the problem of finding all possible GS configurations on the M\"{o}bius strip to finding the GS as a function of $L$. 
The GS configurations, and especially the distribution of the two domains, are visualized in the lower four panels of Fig. \ref{Fig:3} for various values of $L$. 
From these four panels, it can be seen that by varying $L$, we are able to tune the size of the two domains - with the angular domain covering the entire surface for large $L$ and the radial domain covering it for small $L$. 
In the following, we will analyze the size change of the domains when $L$ is increased. 
While we are focusing on a specific example system, the shown behavior, i.e., the evolution of the domains when $L$ is varied, is general and occurs almost exactly the same way regardless of the specific values of $R$ and $N/M$. 
However, before we analyze the domain size, a final comment on the impact of the ratio $N/M$ is in order: 
The overall impact of varying $N/M$ can be described as follows: 
For certain values of $N/M$ (especially towards the extremes $N\gg M$ or $N\ll M$) it may be that (depending on the case) not all of the shown configurations are accessible - unless we make $L$ so large that the surface intersects itself. 
Furthermore, in those regimes where the surface intersects itself, significantly different GS configurations can be found. 
However, the regime where the surface intersects itself is outside of the scope of this work.

To get an idea of how the domain-sizes change with $L$, we classify the ground states by the area $A_\phi\in[0,2\pi L]$ occupied by the angular domain. 
Note that $A_\phi$ refers to the area in parametric coordinates. 
For $A_\phi = 0$, the entire strip is occupied by the radial domain, whereas for $A_\phi = 2\pi L$ the entire strip is covered by the angular domain. 
The value of $A_\phi$ can be obtained for any GS configuration by simply counting all dipoles that significantly align along $\textbf{e}_\phi$. 
The area $A_r$ occupied by the radial domain is then $A_r = 2\pi L - A_\phi$. 
Simulation results of the area $A_\phi$ occupied by the radial domain as a function of $L$ are presented in Fig. \ref{Fig:3} for numerically determined ground states on a M\"{o}bius strip with $N=101$, $M=11$, and $R=1$. 
Note that the data shown in Fig. \ref{Fig:3} are obtained using a NN approximation. 
In the figure, four clearly distinct regions can be seen: 
In region I, only the radial domain exists. 
In region II, both the radial and the angular domain coexist, with the angular domain increasing in size with increasing $L$. 
When the angular domain first appears (the border between regions I and II in Fig. \ref{Fig:3}), it emerges from the point $(\phi,v)=(-\pi/N,-L/2)$. 
From there, the angular domain will mainly grow along $\textbf{e}_\phi$ as $L$ is increased. 
As the angular domain grows, it will eventually have circled around the M\"{o}bius strip and for $L=2 M R \sin(\pi/N)$ connect with itself at the point $(\phi,v) = (\pi-\pi/N,0)$. 
With further increasing $L$, the angular domain will eventually occupy half of the total area of the strip (the border between regions II and III in Fig. \ref{Fig:3}). 
Around the border between regions II and III, the domain sizes change very rapidly - indicating a great sensitivity of this point to parameter variations. 
In region III, the angular domain increases further in size when $L$ is increased, until finally in region IV it encompasses the entire strip. 
For systems with different $N/M$, no discernible differences from the above-described behavior could be observed.

\section{Compression induced topological crossover}\label{Sec:4_crossover}

We will now demonstrate that the GS dipole configuration passes through two distinct crossovers if the length $L$ is varied. 
The variation of $L$ discussed above can also be interpreted as an adiabatic compression or stretching of the M\"{o}bius strip. 
The behavior of the lattice during such a compression can be analyzed with the 2D compression modulus or, simply, 2D modulus \cite{liu2016} of the strip. 
The 2D modulus is defined as
\begin{equation}\label{Eq:2D_modulus}
K = A_0 \frac{d^2 U}{dA^2} = \frac{L_0}{2\pi R} \frac{d^2 U}{dL^2},
\end{equation}
where $A$ is the area of the strip, $A_0$ and $L_0$ denote the area and strip width before compression, and $U$ is the total energy of the system. 
The 2D modulus describes how a change in the width changes the force that is required to compress (or stretch) the strip. 
Consequently, small values of $K$ indicate that the required force changes very little when $L$ is varied, whereas large values of $K$ imply large changes in the required forces when $L$ is varied.

\begin{figure}
	\includegraphics[width=\columnwidth]{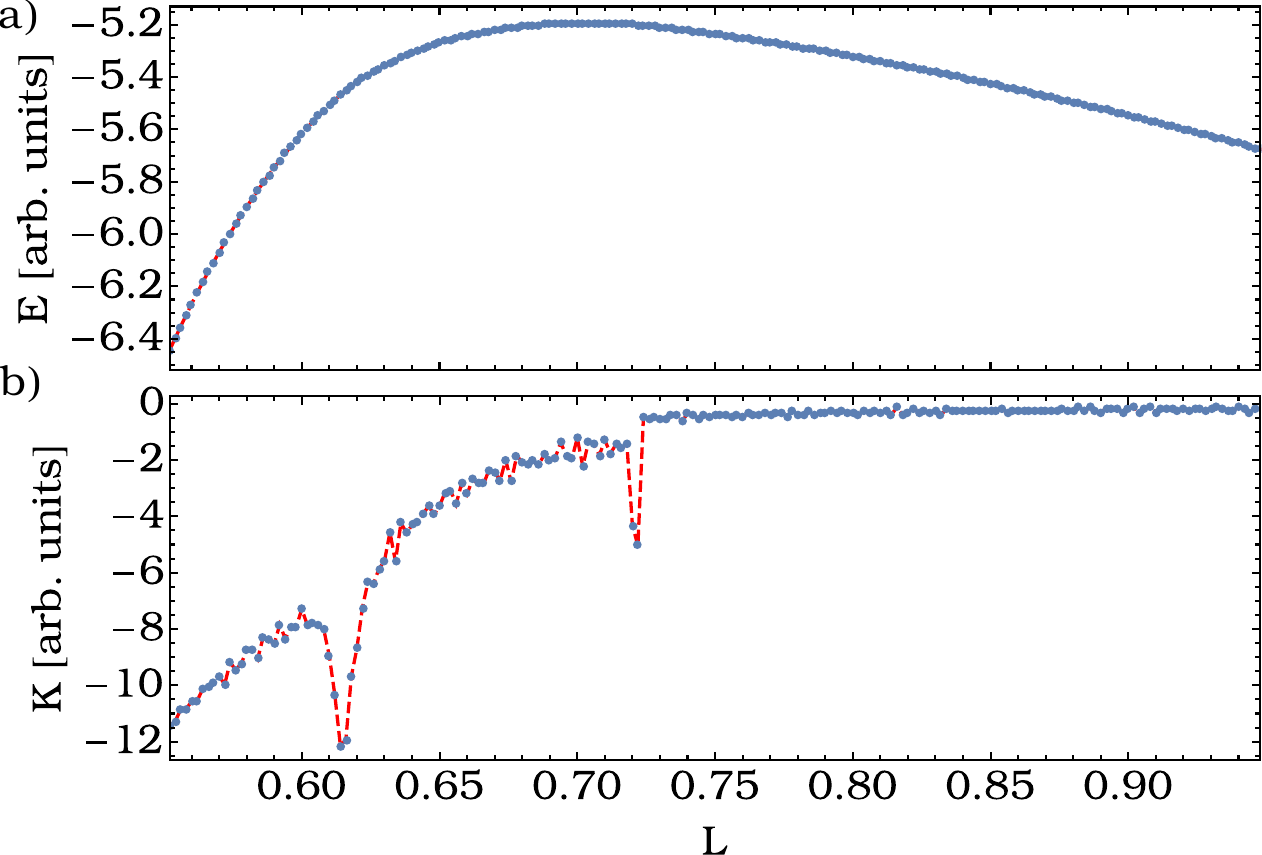}
	\caption{\label{Fig:4} (a) Total energy $U$ as a function of the strip width $L$. (b) The 2D modulus $K\sim d^2U/dL^2$ as a function of the strip width $L$. Both the energy and the 2D modulus data were obtained for a M\"{o}bius strip with $N=101$, $M=11$, and $R=1$. }
\end{figure}

The 2D modulus of Eq. (\ref{Eq:2D_modulus}) depends mainly on the behavior of the total ground state energy $U(L)$ of the system. 
This total energy $U(L)$ as a function of the strip width $L$ is shown in Fig. \ref{Fig:4}(a). 
Interestingly, the curve $U(L)$ exhibits a global maximum. 
We will from now on use $L_{crit}$ to refer to the width of the M\"{o}bius strip at this maximum. 
For values $L<L_{crit}$ the energy increases with increasing strip width, implying that the strip prefers a compressed state and resists stretching. 
On the other hand, for $L>L_{crit}$, the energy decreases with increasing $L$, implying that in this regime the compression of the strip requires energy. 
This crossover from favoring compression to favoring stretching can be understood from the dipole alignments in the radial and angular domains. 
It is a result of the competition between the angular domain favoring stretching and the radial domain favoring compression. 
Within the radial domain, the dipoles are aligned along $\textbf{e}_v$ and will naturally prefer the distance $a$ to their nearest neighbors along the $\textbf{e}_v$-direction to be as small as possible. 
Minimizing $a$ can be achieved globally by decreasing $L$. 
Consequently, decreasing $L$ will decrease the total energy proportional to the number of dipoles in the radial domain. 
On the other hand, in the angular domain, dipoles are aligned along $\textbf{e}_\phi$. 
Consequently, these dipoles will prefer a decrease in the nearest neighbor distance $b(\phi+\Delta \phi,v)$ along the $\textbf{e}_\phi$-direction. 
However, the distance $b$ can change, depending on the position on the strip. 
Furthermore, when $L$ is varied, the distance $b$ can increase in some parts of the strip and decrease in others. 
However, due to the strong decay of the dipole-dipole interactions, the impact of a change in $L$ on the energy is larger for those dipoles where $b$ is smaller. 
The distance $b(\phi+\Delta \phi,v)$ is minimized at $(\phi,v) = (-\pi/N,-L/2)$, i.e. at the origin of the radial domain. 
And around this point, $b$ will decrease further when $L$ is increased.

Close to the crossover point $L_{crit}$, the 2D modulus varies rapidly and has a local minimum [see Fig. \ref{Fig:4}(b)]. 
This peak in the 2D modulus, however, corresponds to the border between regions III and IV in Fig. \ref{Fig:3}. 
For different $N$, $M$ and $R$, this peak does not necessarily coincide with the maximum of $U(L)$. 
Interestingly, a second peak can be seen in the 2D modulus - indicating a second crossover. 
This second peak is appears at a width of $L = 2 M R \sin(\pi/N)$ - very close to the transition between regimes II and III. 
Although this second peak has no discernible effect on the total energy of the system, it does mark a significant change in the structure of the domains. 
For values of $L < 2 M R \sin(\pi/N)$ slightly below the peak, the radial domain not only covers the majority of the M\"{o}bius strip, but also winds around it once - giving the domain a non-trivial topology. 
At the same time, the angular domain has a trivial topology  for $L < 2 M R \sin(\pi/N)$. 
In contrast, for $L > 2 M R \sin(\pi/N)$ above the peak, it is the angular domain that has a non-trivial topology and the radial domain being topologically trivial. 
In summary, the domain which (azimuthally) extends over the complete M\"{o}bius strip inherits its non-trivial topology, whereas domains covering only a finite azimuthal part of the M\"{o}bius strip are topologically trivial.

As described above, during the crossover - as $L$ is increased - the angular domain grows and connects with itself at the point $(\phi,v) = (\pi-\pi/N,0)$. 
To better understand this crossover, it is helpful to analyze $\gamma$ in the vicinity of this point. 
First, we find
\begin{equation}
\frac{d\gamma(\phi,v)}{d\phi}\bigg|_{\phi\rightarrow\pi-\pi/N,v\rightarrow 0} = \frac{d\gamma(\phi,v)}{dv}\bigg|_{\phi\rightarrow\pi-\pi/N,v\rightarrow 0} = 0
\end{equation}
indicating that $\gamma$ always has a critical point at $(\phi,v) = (\pi-\pi/N,0)$. 
Furthermore, at $(\phi,v) = (\pi-\pi/N,0)$ the Hessian matrix of second derivatives is indefinite, indicating that the critical point is a saddle point. 
Exactly for $L = 2 M R \sin(\pi/N)$ the value of $\gamma$ at the saddle point becomes $\gamma(\pi-\pi/N,0) = 1$. 
Any small change of $L$ will lift (or lower) $\gamma$ in the vicinity of the saddle point. 
This is why for $L<2 M R \sin(\pi/N)$ it is the radial domain that winds around the M\"{o}bius strip and for $L>2 M R \sin(\pi/N)$ it is the angular domain. 
For $L=2 M R \sin(\pi/N)$, the system reaches a transition point where neither of the two domains winds around the M\"{o}bius strip. 
For $L=2 M R \sin(\pi/N)$, both domains are topologically trivial; whereas for $L\neq 2 M R \sin(\pi/N)$, one of them is not.

\section{summary and conclusions}\label{Sec:5_end}

We investigated a lattice of interacting dipoles that is spanned on the surface of a M\"{o}bius strip. 
The curved geometry of the lattice can - already for the ground state - lead to the presence of domains with different dipole orientations, as well as domain walls separating these domains. 
Specifically, the GS of the M\"{o}bius strip contains up to two domains - referred to as the angular and the radial domain respectively. 
We discussed the dependence of the ground state on the system parameters and subsequently analyzed the dependence of the GS on the width $L$ of the M\"{o}bius strip.
We demonstrated that by varying $L$, we are able to tune between different GS configurations. 
For large and small $L$, the entire M\"{o}bius strip is exclusively covered by either the radial or the angular domain. 
For intermediate $L$, both domains can be present simultaneously. 
In this intermediate regime, the relative size of the domains can be tuned by varying $L$. 
For a lattice with dimensions $N\cross M$ and a given M\"{o}bius strip radius $R$, one notable GS configuration is reached for $L = 2 M R \sin(\pi/N)$. 
Any increase or decrease of the width from this value will lead to a drastic change in the domain topology. 
Specifically, for wider strips ($L > 2 M R \sin(\pi/N)$), the angular domain will wind around the entire strip, thereby being topologically non-trivial. 
And for narrower strips ($L < 2 M R \sin(\pi/N)$), it is the radial domain that winds around the entire strip. 
We explained this behavior with the presence of a saddle point structure in the $\gamma$ parameter which classifies the domain structure of the entire M\"{o}bius strip. 
Furthermore, we showed that this crossover in the domain topology can be detected as a peak in the 2D compression modulus. 
Additionally, the rapid change in the domain sizes that accompanies this crossover highlights the sensitivity of the crossover point to possible variations of the M\"{o}bius geometry.

In addition to this topological crossover, we also detected a second crossover that can be tuned by varying the width $L$ of the M\"{o}bius strip. 
The crossover point coincides with the maximum of the total (width-dependent) ground state energy $U(L)$. 
During the crossover, the system changes from a state that resists compression to a state that favors compression. 
Consequently, this crossover point will be quite sensitive to variations of the M\"{o}bius strip width $L$ - and by extension to variations of the system parameters $R$, $N$ and $M$. 

While our analysis of the dipolar lattice on a M\"{o}bius strip shows already an intricate structure formation for the ground state there is several open future directions of research. 
An immediate case of investigation would be the low-lying excited states and their properties. 
How and where do topological and non-topological defects and possibly kinks emerge in the  dipolar lattice and how do they 'interact' with the domain walls? 
Quenches of the geometrical parameters across the phase boundaries would be promising candidates for dynamical and transient structure formation in the higher energy regime. 
As a longer term and promising perspective we envisage the investigation of dipolar lattices on geometrically and topologically more complex curved surfaces. 
While there is a plethora of such surfaces in particular in the framework of (multiply-periodic) minimal surfaces \cite{dierkes2010} the impact of the dipolar interaction on self-intersecting surfaces is an open and intriguing problem to be explored in the future.

A final remark concerning the experimental preparation of such surfaces is in order.
Curvilinear flat architectures can be prepared in the framework of nanostructures using conventional techniques based on thin film deposition and lithographic methods \cite{sheka2021}. 
Ion-beam writing techniques represent another alternative. 
Fabrication of complex 3D nanoarchitectures is challenging and requires even more advanced and specialized preparation tools in particular if it comes to the combination with magnetic sublattices \cite{sheka2021}. 
As a conclusion we remark that while the experimental implementation of the dipolar lattices on curved surfaces is certainly highly demanding the richness of their phenomenology and perspectives render them highly promising candidates for future investigation.

\section*{Acknowledgments}

F.A.O.S acknowledges CAPES (No. 88887.670331/2022-00) for financial support. 

\bibliography{txtest}

\begin{thebibliography}{36}%
\makeatletter
\providecommand \@ifxundefined [1]{%
 \@ifx{#1\undefined}
}%
\providecommand \@ifnum [1]{%
 \ifnum #1\expandafter \@firstoftwo
 \else \expandafter \@secondoftwo
 \fi
}%
\providecommand \@ifx [1]{%
 \ifx #1\expandafter \@firstoftwo
 \else \expandafter \@secondoftwo
 \fi
}%
\providecommand \natexlab [1]{#1}%
\providecommand \enquote  [1]{``#1''}%
\providecommand \bibnamefont  [1]{#1}%
\providecommand \bibfnamefont [1]{#1}%
\providecommand \citenamefont [1]{#1}%
\providecommand \href@noop [0]{\@secondoftwo}%
\providecommand \href [0]{\begingroup \@sanitize@url \@href}%
\providecommand \@href[1]{\@@startlink{#1}\@@href}%
\providecommand \@@href[1]{\endgroup#1\@@endlink}%
\providecommand \@sanitize@url [0]{\catcode `\\12\catcode `\$12\catcode
  `\&12\catcode `\#12\catcode `\^12\catcode `\_12\catcode `\%12\relax}%
\providecommand \@@startlink[1]{}%
\providecommand \@@endlink[0]{}%
\providecommand \url  [0]{\begingroup\@sanitize@url \@url }%
\providecommand \@url [1]{\endgroup\@href {#1}{\urlprefix }}%
\providecommand \urlprefix  [0]{URL }%
\providecommand \Eprint [0]{\href }%
\providecommand \doibase [0]{http://dx.doi.org/}%
\providecommand \selectlanguage [0]{\@gobble}%
\providecommand \bibinfo  [0]{\@secondoftwo}%
\providecommand \bibfield  [0]{\@secondoftwo}%
\providecommand \translation [1]{[#1]}%
\providecommand \BibitemOpen [0]{}%
\providecommand \bibitemStop [0]{}%
\providecommand \bibitemNoStop [0]{.\EOS\space}%
\providecommand \EOS [0]{\spacefactor3000\relax}%
\providecommand \BibitemShut  [1]{\csname bibitem#1\endcsname}%
\let\auto@bib@innerbib\@empty
\bibitem [{\citenamefont {Samajdar}\ \emph {et~al.}(2020)\citenamefont
  {Samajdar}, \citenamefont {Ho}, \citenamefont {Pichler}, \citenamefont
  {Lukin},\ and\ \citenamefont {Sachdev}}]{samajdar2020}%
  \BibitemOpen
  \bibfield  {author} {\bibinfo {author} {\bibfnamefont {R.}~\bibnamefont
  {Samajdar}}, \bibinfo {author} {\bibfnamefont {W.~W.}\ \bibnamefont {Ho}},
  \bibinfo {author} {\bibfnamefont {H.}~\bibnamefont {Pichler}}, \bibinfo
  {author} {\bibfnamefont {M.~D.}\ \bibnamefont {Lukin}}, \ and\ \bibinfo
  {author} {\bibfnamefont {S.}~\bibnamefont {Sachdev}},\ }\href {\doibase
  10.1103/PhysRevLett.124.103601} {\bibfield  {journal} {\bibinfo  {journal}
  {Phys. Rev. Lett.}\ }\textbf {\bibinfo {volume} {124}},\ \bibinfo {pages}
  {103601} (\bibinfo {year} {2020})}\BibitemShut {NoStop}%
\bibitem [{\citenamefont {Chomaz}\ \emph {et~al.}(2023)\citenamefont {Chomaz},
  \citenamefont {{Ferrier-Barbut}}, \citenamefont {Ferlaino}, \citenamefont
  {{Laburthe-Tolra}}, \citenamefont {Lev},\ and\ \citenamefont
  {Pfau}}]{chomaz2023}%
  \BibitemOpen
  \bibfield  {author} {\bibinfo {author} {\bibfnamefont {L.}~\bibnamefont
  {Chomaz}}, \bibinfo {author} {\bibfnamefont {I.}~\bibnamefont
  {{Ferrier-Barbut}}}, \bibinfo {author} {\bibfnamefont {F.}~\bibnamefont
  {Ferlaino}}, \bibinfo {author} {\bibfnamefont {B.}~\bibnamefont
  {{Laburthe-Tolra}}}, \bibinfo {author} {\bibfnamefont {B.~L.}\ \bibnamefont
  {Lev}}, \ and\ \bibinfo {author} {\bibfnamefont {T.}~\bibnamefont {Pfau}},\
  }\href {\doibase 10.1088/1361-6633/aca814} {\bibfield  {journal} {\bibinfo
  {journal} {Rep. Prog. Phys.}\ }\textbf {\bibinfo {volume} {86}},\ \bibinfo
  {pages} {026401} (\bibinfo {year} {2023})}\BibitemShut {NoStop}%
\bibitem [{\citenamefont {Ruhman}\ \emph {et~al.}(2012)\citenamefont {Ruhman},
  \citenamefont {Dalla~Torre}, \citenamefont {Huber},\ and\ \citenamefont
  {Altman}}]{ruhman2012}%
  \BibitemOpen
  \bibfield  {author} {\bibinfo {author} {\bibfnamefont {J.}~\bibnamefont
  {Ruhman}}, \bibinfo {author} {\bibfnamefont {E.~G.}\ \bibnamefont
  {Dalla~Torre}}, \bibinfo {author} {\bibfnamefont {S.~D.}\ \bibnamefont
  {Huber}}, \ and\ \bibinfo {author} {\bibfnamefont {E.}~\bibnamefont
  {Altman}},\ }\href {\doibase 10.1103/PhysRevB.85.125121} {\bibfield
  {journal} {\bibinfo  {journal} {Phys. Rev. B}\ }\textbf {\bibinfo {volume}
  {85}},\ \bibinfo {pages} {125121} (\bibinfo {year} {2012})}\BibitemShut
  {NoStop}%
\bibitem [{\citenamefont {Johnston}(2016)}]{johnston2016}%
  \BibitemOpen
  \bibfield  {author} {\bibinfo {author} {\bibfnamefont {D.~C.}\ \bibnamefont
  {Johnston}},\ }\href {\doibase 10.1103/PhysRevB.93.014421} {\bibfield
  {journal} {\bibinfo  {journal} {Phys. Rev. B}\ }\textbf {\bibinfo {volume}
  {93}},\ \bibinfo {pages} {014421} (\bibinfo {year} {2016})}\BibitemShut
  {NoStop}%
\bibitem [{\citenamefont {Qiao}\ \emph {et~al.}(2021)\citenamefont {Qiao},
  \citenamefont {Wang}, \citenamefont {Choi}, \citenamefont {Park},\ and\
  \citenamefont {Kim}}]{qiao2021}%
  \BibitemOpen
  \bibfield  {author} {\bibinfo {author} {\bibfnamefont {H.}~\bibnamefont
  {Qiao}}, \bibinfo {author} {\bibfnamefont {C.}~\bibnamefont {Wang}}, \bibinfo
  {author} {\bibfnamefont {W.~S.}\ \bibnamefont {Choi}}, \bibinfo {author}
  {\bibfnamefont {M.~H.}\ \bibnamefont {Park}}, \ and\ \bibinfo {author}
  {\bibfnamefont {Y.}~\bibnamefont {Kim}},\ }\href {\doibase
  10.1016/j.mser.2021.100622} {\bibfield  {journal} {\bibinfo  {journal}
  {Mater. Sci. Eng. R-Rep.}\ }\textbf {\bibinfo {volume} {145}},\ \bibinfo
  {pages} {100622} (\bibinfo {year} {2021})}\BibitemShut {NoStop}%
\bibitem [{\citenamefont {K{\"a}nzig}(1957)}]{kanzig1957}%
  \BibitemOpen
  \bibfield  {author} {\bibinfo {author} {\bibfnamefont {W.}~\bibnamefont
  {K{\"a}nzig}},\ }in\ \href {\doibase 10.1016/S0081-1947(08)60154-X} {\emph
  {\bibinfo {booktitle} {Solid {{State Physics}}}}},\ Vol.~\bibinfo {volume}
  {4}\ (\bibinfo  {publisher} {{Elsevier}},\ \bibinfo {year} {1957})\ pp.\
  \bibinfo {pages} {1--197}\BibitemShut {NoStop}%
\bibitem [{\citenamefont {Martin}\ and\ \citenamefont
  {Rappe}(2017)}]{martin2017}%
  \BibitemOpen
  \bibfield  {author} {\bibinfo {author} {\bibfnamefont {L.~W.}\ \bibnamefont
  {Martin}}\ and\ \bibinfo {author} {\bibfnamefont {A.~M.}\ \bibnamefont
  {Rappe}},\ }\href {\doibase 10.1038/natrevmats.2016.87} {\bibfield  {journal}
  {\bibinfo  {journal} {Nat Rev Mater}\ }\textbf {\bibinfo {volume} {2}},\
  \bibinfo {pages} {16087} (\bibinfo {year} {2017})}\BibitemShut {NoStop}%
\bibitem [{\citenamefont {Meier}\ and\ \citenamefont
  {Selbach}(2022)}]{meier2022}%
  \BibitemOpen
  \bibfield  {author} {\bibinfo {author} {\bibfnamefont {D.}~\bibnamefont
  {Meier}}\ and\ \bibinfo {author} {\bibfnamefont {S.~M.}\ \bibnamefont
  {Selbach}},\ }\href {\doibase 10.1038/s41578-021-00375-z} {\bibfield
  {journal} {\bibinfo  {journal} {Nat. Rev. Mater.}\ }\textbf {\bibinfo
  {volume} {7}},\ \bibinfo {pages} {157} (\bibinfo {year} {2022})}\BibitemShut
  {NoStop}%
\bibitem [{\citenamefont {Sharma}\ \emph {et~al.}(2017)\citenamefont {Sharma},
  \citenamefont {Zhang}, \citenamefont {Sando}, \citenamefont {Lei},
  \citenamefont {Liu}, \citenamefont {Li}, \citenamefont {Nagarajan},\ and\
  \citenamefont {Seidel}}]{sharma2017}%
  \BibitemOpen
  \bibfield  {author} {\bibinfo {author} {\bibfnamefont {P.}~\bibnamefont
  {Sharma}}, \bibinfo {author} {\bibfnamefont {Q.}~\bibnamefont {Zhang}},
  \bibinfo {author} {\bibfnamefont {D.}~\bibnamefont {Sando}}, \bibinfo
  {author} {\bibfnamefont {C.~H.}\ \bibnamefont {Lei}}, \bibinfo {author}
  {\bibfnamefont {Y.}~\bibnamefont {Liu}}, \bibinfo {author} {\bibfnamefont
  {J.}~\bibnamefont {Li}}, \bibinfo {author} {\bibfnamefont {V.}~\bibnamefont
  {Nagarajan}}, \ and\ \bibinfo {author} {\bibfnamefont {J.}~\bibnamefont
  {Seidel}},\ }\href {\doibase 10.1126/sciadv.1700512} {\bibfield  {journal}
  {\bibinfo  {journal} {Sci. Adv.}\ }\textbf {\bibinfo {volume} {3}},\ \bibinfo
  {pages} {e1700512} (\bibinfo {year} {2017})}\BibitemShut {NoStop}%
\bibitem [{\citenamefont {Wang}\ \emph {et~al.}(2022)\citenamefont {Wang},
  \citenamefont {Ma}, \citenamefont {Huang}, \citenamefont {Ma}, \citenamefont
  {Jafri}, \citenamefont {Fan}, \citenamefont {Yang}, \citenamefont {Wang},
  \citenamefont {Chen}, \citenamefont {Liu}, \citenamefont {Zhang},
  \citenamefont {Lin}, \citenamefont {Chen}, \citenamefont {Yi},\ and\
  \citenamefont {Nan}}]{wang2022}%
  \BibitemOpen
  \bibfield  {author} {\bibinfo {author} {\bibfnamefont {J.}~\bibnamefont
  {Wang}}, \bibinfo {author} {\bibfnamefont {J.}~\bibnamefont {Ma}}, \bibinfo
  {author} {\bibfnamefont {H.}~\bibnamefont {Huang}}, \bibinfo {author}
  {\bibfnamefont {J.}~\bibnamefont {Ma}}, \bibinfo {author} {\bibfnamefont
  {H.~M.}\ \bibnamefont {Jafri}}, \bibinfo {author} {\bibfnamefont
  {Y.}~\bibnamefont {Fan}}, \bibinfo {author} {\bibfnamefont {H.}~\bibnamefont
  {Yang}}, \bibinfo {author} {\bibfnamefont {Y.}~\bibnamefont {Wang}}, \bibinfo
  {author} {\bibfnamefont {M.}~\bibnamefont {Chen}}, \bibinfo {author}
  {\bibfnamefont {D.}~\bibnamefont {Liu}}, \bibinfo {author} {\bibfnamefont
  {J.}~\bibnamefont {Zhang}}, \bibinfo {author} {\bibfnamefont {Y.-H.}\
  \bibnamefont {Lin}}, \bibinfo {author} {\bibfnamefont {L.-Q.}\ \bibnamefont
  {Chen}}, \bibinfo {author} {\bibfnamefont {D.}~\bibnamefont {Yi}}, \ and\
  \bibinfo {author} {\bibfnamefont {C.-W.}\ \bibnamefont {Nan}},\ }\href
  {\doibase 10.1038/s41467-022-30983-4} {\bibfield  {journal} {\bibinfo
  {journal} {Nat. Commun.}\ }\textbf {\bibinfo {volume} {13}},\ \bibinfo
  {pages} {3255} (\bibinfo {year} {2022})}\BibitemShut {NoStop}%
\bibitem [{\citenamefont {Muralt}(2000)}]{muralt2000}%
  \BibitemOpen
  \bibfield  {author} {\bibinfo {author} {\bibfnamefont {P.}~\bibnamefont
  {Muralt}},\ }\href {\doibase 10.1088/0960-1317/10/2/307} {\bibfield
  {journal} {\bibinfo  {journal} {J. Micromech. Microeng.}\ }\textbf {\bibinfo
  {volume} {10}},\ \bibinfo {pages} {136} (\bibinfo {year} {2000})}\BibitemShut
  {NoStop}%
\bibitem [{\citenamefont {Schildknecht}\ \emph {et~al.}(2019)\citenamefont
  {Schildknecht}, \citenamefont {Sch{\"u}tt}, \citenamefont {Heyderman},\ and\
  \citenamefont {Derlet}}]{schildknecht2019}%
  \BibitemOpen
  \bibfield  {author} {\bibinfo {author} {\bibfnamefont {D.}~\bibnamefont
  {Schildknecht}}, \bibinfo {author} {\bibfnamefont {M.}~\bibnamefont
  {Sch{\"u}tt}}, \bibinfo {author} {\bibfnamefont {L.~J.}\ \bibnamefont
  {Heyderman}}, \ and\ \bibinfo {author} {\bibfnamefont {P.~M.}\ \bibnamefont
  {Derlet}},\ }\href {\doibase 10.1103/PhysRevB.100.014426} {\bibfield
  {journal} {\bibinfo  {journal} {Phys. Rev. B}\ }\textbf {\bibinfo {volume}
  {100}},\ \bibinfo {pages} {014426} (\bibinfo {year} {2019})}\BibitemShut
  {NoStop}%
\bibitem [{\citenamefont {Brankov}\ and\ \citenamefont
  {Danchev}(1987)}]{brankov1987}%
  \BibitemOpen
  \bibfield  {author} {\bibinfo {author} {\bibfnamefont {J.}~\bibnamefont
  {Brankov}}\ and\ \bibinfo {author} {\bibfnamefont {D.}~\bibnamefont
  {Danchev}},\ }\href {\doibase 10.1016/0378-4371(87)90148-8} {\bibfield
  {journal} {\bibinfo  {journal} {Physica A}\ }\textbf {\bibinfo {volume}
  {144}},\ \bibinfo {pages} {128} (\bibinfo {year} {1987})}\BibitemShut
  {NoStop}%
\bibitem [{\citenamefont {De'Bell}\ \emph {et~al.}(1997)\citenamefont
  {De'Bell}, \citenamefont {MacIsaac}, \citenamefont {Booth},\ and\
  \citenamefont {Whitehead}}]{debell1997}%
  \BibitemOpen
  \bibfield  {author} {\bibinfo {author} {\bibfnamefont {K.}~\bibnamefont
  {De'Bell}}, \bibinfo {author} {\bibfnamefont {A.~B.}\ \bibnamefont
  {MacIsaac}}, \bibinfo {author} {\bibfnamefont {I.~N.}\ \bibnamefont {Booth}},
  \ and\ \bibinfo {author} {\bibfnamefont {J.~P.}\ \bibnamefont {Whitehead}},\
  }\href {\doibase 10.1103/PhysRevB.55.15108} {\bibfield  {journal} {\bibinfo
  {journal} {Phys. Rev. B}\ }\textbf {\bibinfo {volume} {55}},\ \bibinfo
  {pages} {15108} (\bibinfo {year} {1997})}\BibitemShut {NoStop}%
\bibitem [{\citenamefont {Feldmann}\ \emph {et~al.}(2008)\citenamefont
  {Feldmann}, \citenamefont {Kalman}, \citenamefont {Hartmann},\ and\
  \citenamefont {Rosenberg}}]{feldmann2008}%
  \BibitemOpen
  \bibfield  {author} {\bibinfo {author} {\bibfnamefont {J.~D.}\ \bibnamefont
  {Feldmann}}, \bibinfo {author} {\bibfnamefont {G.~J.}\ \bibnamefont
  {Kalman}}, \bibinfo {author} {\bibfnamefont {P.}~\bibnamefont {Hartmann}}, \
  and\ \bibinfo {author} {\bibfnamefont {M.}~\bibnamefont {Rosenberg}},\ }\href
  {\doibase 10.1103/PhysRevLett.100.085001} {\bibfield  {journal} {\bibinfo
  {journal} {Phys. Rev. Lett.}\ }\textbf {\bibinfo {volume} {100}},\ \bibinfo
  {pages} {085001} (\bibinfo {year} {2008})}\BibitemShut {NoStop}%
\bibitem [{\citenamefont {Mydosh}(1996)}]{mydosh1996}%
  \BibitemOpen
  \bibfield  {author} {\bibinfo {author} {\bibfnamefont {J.}~\bibnamefont
  {Mydosh}},\ }\href {\doibase 10.1016/0304-8853(95)01272-9} {\bibfield
  {journal} {\bibinfo  {journal} {Journal of Magnetism and Magnetic Materials}\
  }\textbf {\bibinfo {volume} {157--158}},\ \bibinfo {pages} {606} (\bibinfo
  {year} {1996})}\BibitemShut {NoStop}%
\bibitem [{\citenamefont {Ramirez}(2001)}]{ramirez2001}%
  \BibitemOpen
  \bibfield  {author} {\bibinfo {author} {\bibfnamefont {A.}~\bibnamefont
  {Ramirez}},\ }in\ \href {\doibase 10.1016/S1567-2719(01)13008-8} {\emph
  {\bibinfo {booktitle} {Handbook of {{Magnetic Materials}}}}},\ Vol.~\bibinfo
  {volume} {13}\ (\bibinfo  {publisher} {{Elsevier}},\ \bibinfo {year} {2001})\
  pp.\ \bibinfo {pages} {423--520}\BibitemShut {NoStop}%
\bibitem [{\citenamefont {Melchy}\ and\ \citenamefont
  {Zhitomirsky}(2009)}]{melchy2009}%
  \BibitemOpen
  \bibfield  {author} {\bibinfo {author} {\bibfnamefont {P.-{\'E}.}\
  \bibnamefont {Melchy}}\ and\ \bibinfo {author} {\bibfnamefont {M.~E.}\
  \bibnamefont {Zhitomirsky}},\ }\href {\doibase 10.1103/PhysRevB.80.064411}
  {\bibfield  {journal} {\bibinfo  {journal} {Phys. Rev. B}\ }\textbf {\bibinfo
  {volume} {80}},\ \bibinfo {pages} {064411} (\bibinfo {year}
  {2009})}\BibitemShut {NoStop}%
\bibitem [{\citenamefont {Gaulin}(1994)}]{gaulin1994}%
  \BibitemOpen
  \bibfield  {author} {\bibinfo {author} {\bibfnamefont {B.~D.}\ \bibnamefont
  {Gaulin}},\ }\href {\doibase 10.1007/BF02069416} {\bibfield  {journal}
  {\bibinfo  {journal} {Hyperfine Interact.}\ }\textbf {\bibinfo {volume}
  {85}},\ \bibinfo {pages} {159} (\bibinfo {year} {1994})}\BibitemShut
  {NoStop}%
\bibitem [{\citenamefont {Siemens}\ and\ \citenamefont
  {Schmelcher}(2022)}]{siemens2022}%
  \BibitemOpen
  \bibfield  {author} {\bibinfo {author} {\bibfnamefont {A.}~\bibnamefont
  {Siemens}}\ and\ \bibinfo {author} {\bibfnamefont {P.}~\bibnamefont
  {Schmelcher}},\ }\href {\doibase 10.1088/1751-8121/ac86af} {\bibfield
  {journal} {\bibinfo  {journal} {J. Phys. A: Math. Theor.}\ }\textbf {\bibinfo
  {volume} {55}},\ \bibinfo {pages} {375205} (\bibinfo {year}
  {2022})}\BibitemShut {NoStop}%
\bibitem [{\citenamefont {Lieu}\ and\ \citenamefont
  {Yoshinaga}(2020)}]{lieu2020}%
  \BibitemOpen
  \bibfield  {author} {\bibinfo {author} {\bibfnamefont {U.~T.}\ \bibnamefont
  {Lieu}}\ and\ \bibinfo {author} {\bibfnamefont {N.}~\bibnamefont
  {Yoshinaga}},\ }\href {\doibase 10.1039/D0SM00103A} {\bibfield  {journal}
  {\bibinfo  {journal} {Soft Matter}\ }\textbf {\bibinfo {volume} {16}},\
  \bibinfo {pages} {7667} (\bibinfo {year} {2020})}\BibitemShut {NoStop}%
\bibitem [{\citenamefont {Siemens}\ and\ \citenamefont
  {Schmelcher}(2023)}]{siemens2023}%
  \BibitemOpen
  \bibfield  {author} {\bibinfo {author} {\bibfnamefont {A.}~\bibnamefont
  {Siemens}}\ and\ \bibinfo {author} {\bibfnamefont {P.}~\bibnamefont
  {Schmelcher}},\ }\href {\doibase 10.1088/1751-8121/ad0bcb} {\bibfield
  {journal} {\bibinfo  {journal} {J. Phys. A: Math. Theor.}\ }\textbf {\bibinfo
  {volume} {56}},\ \bibinfo {pages} {495702} (\bibinfo {year}
  {2023})}\BibitemShut {NoStop}%
\bibitem [{\citenamefont {Schmelcher}(2011)}]{schmelcher2011}%
  \BibitemOpen
  \bibfield  {author} {\bibinfo {author} {\bibfnamefont {P.}~\bibnamefont
  {Schmelcher}},\ }\href {\doibase 10.1209/0295-5075/95/50005} {\bibfield
  {journal} {\bibinfo  {journal} {Europhys. Lett.}\ }\textbf {\bibinfo {volume}
  {95}},\ \bibinfo {pages} {50005} (\bibinfo {year} {2011})}\BibitemShut
  {NoStop}%
\bibitem [{\citenamefont {Plettenberg}\ \emph {et~al.}(2017)\citenamefont
  {Plettenberg}, \citenamefont {Stockhofe}, \citenamefont {Zampetaki},\ and\
  \citenamefont {Schmelcher}}]{plettenberg2017}%
  \BibitemOpen
  \bibfield  {author} {\bibinfo {author} {\bibfnamefont {J.}~\bibnamefont
  {Plettenberg}}, \bibinfo {author} {\bibfnamefont {J.}~\bibnamefont
  {Stockhofe}}, \bibinfo {author} {\bibfnamefont {A.~V.}\ \bibnamefont
  {Zampetaki}}, \ and\ \bibinfo {author} {\bibfnamefont {P.}~\bibnamefont
  {Schmelcher}},\ }\href {\doibase 10.1103/PhysRevE.95.012213} {\bibfield
  {journal} {\bibinfo  {journal} {Phys. Rev. E}\ }\textbf {\bibinfo {volume}
  {95}},\ \bibinfo {pages} {012213} (\bibinfo {year} {2017})}\BibitemShut
  {NoStop}%
\bibitem [{\citenamefont {Zampetaki}\ \emph {et~al.}(2018)\citenamefont
  {Zampetaki}, \citenamefont {Stockhofe},\ and\ \citenamefont
  {Schmelcher}}]{zampetaki2018}%
  \BibitemOpen
  \bibfield  {author} {\bibinfo {author} {\bibfnamefont {A.~V.}\ \bibnamefont
  {Zampetaki}}, \bibinfo {author} {\bibfnamefont {J.}~\bibnamefont
  {Stockhofe}}, \ and\ \bibinfo {author} {\bibfnamefont {P.}~\bibnamefont
  {Schmelcher}},\ }\href {\doibase 10.1103/PhysRevE.97.042503} {\bibfield
  {journal} {\bibinfo  {journal} {Phys. Rev. E}\ }\textbf {\bibinfo {volume}
  {97}},\ \bibinfo {pages} {042503} (\bibinfo {year} {2018})}\BibitemShut
  {NoStop}%
\bibitem [{\citenamefont {Siemens}\ and\ \citenamefont
  {Schmelcher}(2020)}]{siemens2020}%
  \BibitemOpen
  \bibfield  {author} {\bibinfo {author} {\bibfnamefont {A.}~\bibnamefont
  {Siemens}}\ and\ \bibinfo {author} {\bibfnamefont {P.}~\bibnamefont
  {Schmelcher}},\ }\href {\doibase 10.1103/PhysRevE.102.012147} {\bibfield
  {journal} {\bibinfo  {journal} {Phys. Rev. E}\ }\textbf {\bibinfo {volume}
  {102}},\ \bibinfo {pages} {012147} (\bibinfo {year} {2020})}\BibitemShut
  {NoStop}%
\bibitem [{\citenamefont {Zampetaki}\ \emph {et~al.}(2013)\citenamefont
  {Zampetaki}, \citenamefont {Stockhofe}, \citenamefont {Kr{\"o}nke},\ and\
  \citenamefont {Schmelcher}}]{zampetaki2013}%
  \BibitemOpen
  \bibfield  {author} {\bibinfo {author} {\bibfnamefont {A.~V.}\ \bibnamefont
  {Zampetaki}}, \bibinfo {author} {\bibfnamefont {J.}~\bibnamefont
  {Stockhofe}}, \bibinfo {author} {\bibfnamefont {S.}~\bibnamefont
  {Kr{\"o}nke}}, \ and\ \bibinfo {author} {\bibfnamefont {P.}~\bibnamefont
  {Schmelcher}},\ }\href {\doibase 10.1103/PhysRevE.88.043202} {\bibfield
  {journal} {\bibinfo  {journal} {Phys. Rev. E}\ }\textbf {\bibinfo {volume}
  {88}},\ \bibinfo {pages} {043202} (\bibinfo {year} {2013})}\BibitemShut
  {NoStop}%
\bibitem [{\citenamefont {Zampetaki}\ \emph
  {et~al.}(2015{\natexlab{a}})\citenamefont {Zampetaki}, \citenamefont
  {Stockhofe},\ and\ \citenamefont {Schmelcher}}]{zampetaki2015}%
  \BibitemOpen
  \bibfield  {author} {\bibinfo {author} {\bibfnamefont {A.~V.}\ \bibnamefont
  {Zampetaki}}, \bibinfo {author} {\bibfnamefont {J.}~\bibnamefont
  {Stockhofe}}, \ and\ \bibinfo {author} {\bibfnamefont {P.}~\bibnamefont
  {Schmelcher}},\ }\href {\doibase 10.1103/PhysRevA.91.023409} {\bibfield
  {journal} {\bibinfo  {journal} {Phys. Rev. A}\ }\textbf {\bibinfo {volume}
  {91}},\ \bibinfo {pages} {023409} (\bibinfo {year}
  {2015}{\natexlab{a}})}\BibitemShut {NoStop}%
\bibitem [{\citenamefont {Zampetaki}\ \emph
  {et~al.}(2015{\natexlab{b}})\citenamefont {Zampetaki}, \citenamefont
  {Stockhofe},\ and\ \citenamefont {Schmelcher}}]{zampetaki2015a}%
  \BibitemOpen
  \bibfield  {author} {\bibinfo {author} {\bibfnamefont {A.~V.}\ \bibnamefont
  {Zampetaki}}, \bibinfo {author} {\bibfnamefont {J.}~\bibnamefont
  {Stockhofe}}, \ and\ \bibinfo {author} {\bibfnamefont {P.}~\bibnamefont
  {Schmelcher}},\ }\href {\doibase 10.1103/PhysRevE.92.042905} {\bibfield
  {journal} {\bibinfo  {journal} {Phys. Rev. E}\ }\textbf {\bibinfo {volume}
  {92}},\ \bibinfo {pages} {042905} (\bibinfo {year}
  {2015}{\natexlab{b}})}\BibitemShut {NoStop}%
\bibitem [{\citenamefont {Zampetaki}\ \emph {et~al.}(2017)\citenamefont
  {Zampetaki}, \citenamefont {Stockhofe},\ and\ \citenamefont
  {Schmelcher}}]{zampetaki2017}%
  \BibitemOpen
  \bibfield  {author} {\bibinfo {author} {\bibfnamefont {A.~V.}\ \bibnamefont
  {Zampetaki}}, \bibinfo {author} {\bibfnamefont {J.}~\bibnamefont
  {Stockhofe}}, \ and\ \bibinfo {author} {\bibfnamefont {P.}~\bibnamefont
  {Schmelcher}},\ }\href {\doibase 10.1103/PhysRevE.95.022205} {\bibfield
  {journal} {\bibinfo  {journal} {Phys. Rev. E}\ }\textbf {\bibinfo {volume}
  {95}},\ \bibinfo {pages} {022205} (\bibinfo {year} {2017})}\BibitemShut
  {NoStop}%
\bibitem [{\citenamefont {Siemens}\ and\ \citenamefont
  {Schmelcher}(2021)}]{siemens2021}%
  \BibitemOpen
  \bibfield  {author} {\bibinfo {author} {\bibfnamefont {A.}~\bibnamefont
  {Siemens}}\ and\ \bibinfo {author} {\bibfnamefont {P.}~\bibnamefont
  {Schmelcher}},\ }\href {\doibase 10.1103/PhysRevE.103.052217} {\bibfield
  {journal} {\bibinfo  {journal} {Phys. Rev. E}\ }\textbf {\bibinfo {volume}
  {103}},\ \bibinfo {pages} {052217} (\bibinfo {year} {2021})}\BibitemShut
  {NoStop}%
\bibitem [{\citenamefont {Beugeling}\ \emph {et~al.}(2014)\citenamefont
  {Beugeling}, \citenamefont {Quelle},\ and\ \citenamefont
  {Morais~Smith}}]{beugeling2014}%
  \BibitemOpen
  \bibfield  {author} {\bibinfo {author} {\bibfnamefont {W.}~\bibnamefont
  {Beugeling}}, \bibinfo {author} {\bibfnamefont {A.}~\bibnamefont {Quelle}}, \
  and\ \bibinfo {author} {\bibfnamefont {C.}~\bibnamefont {Morais~Smith}},\
  }\href {\doibase 10.1103/PhysRevB.89.235112} {\bibfield  {journal} {\bibinfo
  {journal} {Phys. Rev. B}\ }\textbf {\bibinfo {volume} {89}},\ \bibinfo
  {pages} {235112} (\bibinfo {year} {2014})}\BibitemShut {NoStop}%
\bibitem [{\citenamefont {Flouris}\ \emph {et~al.}(2022)\citenamefont
  {Flouris}, \citenamefont {Jimenez},\ and\ \citenamefont
  {Herrmann}}]{flouris2022}%
  \BibitemOpen
  \bibfield  {author} {\bibinfo {author} {\bibfnamefont {K.}~\bibnamefont
  {Flouris}}, \bibinfo {author} {\bibfnamefont {M.~M.}\ \bibnamefont
  {Jimenez}}, \ and\ \bibinfo {author} {\bibfnamefont {H.~J.}\ \bibnamefont
  {Herrmann}},\ }\href {\doibase 10.1103/PhysRevB.105.235122} {\bibfield
  {journal} {\bibinfo  {journal} {Phys. Rev. B}\ }\textbf {\bibinfo {volume}
  {105}},\ \bibinfo {pages} {235122} (\bibinfo {year} {2022})}\BibitemShut
  {NoStop}%
\bibitem [{\citenamefont {Liu}\ and\ \citenamefont {Wu}(2016)}]{liu2016}%
  \BibitemOpen
  \bibfield  {author} {\bibinfo {author} {\bibfnamefont {K.}~\bibnamefont
  {Liu}}\ and\ \bibinfo {author} {\bibfnamefont {J.}~\bibnamefont {Wu}},\
  }\href {\doibase 10.1557/jmr.2015.324} {\bibfield  {journal} {\bibinfo
  {journal} {J. Mater. Res.}\ }\textbf {\bibinfo {volume} {31}},\ \bibinfo
  {pages} {832} (\bibinfo {year} {2016})}\BibitemShut {NoStop}%
\bibitem [{\citenamefont {Dierkes}\ \emph {et~al.}(2010)\citenamefont
  {Dierkes}, \citenamefont {Hildebrandt},\ and\ \citenamefont
  {Sauvigny}}]{dierkes2010}%
  \BibitemOpen
  \bibfield  {author} {\bibinfo {author} {\bibfnamefont {U.}~\bibnamefont
  {Dierkes}}, \bibinfo {author} {\bibfnamefont {S.}~\bibnamefont
  {Hildebrandt}}, \ and\ \bibinfo {author} {\bibfnamefont {F.}~\bibnamefont
  {Sauvigny}},\ }\href {\doibase 10.1007/978-3-642-11698-8} {\emph {\bibinfo
  {title} {Minimal {{Surfaces}}}}},\ \bibinfo {series} {A {{Series}} of
  {{Comprehensive Studies}} in {{Mathematics}}}, Vol.\ \bibinfo {volume} {339}\
  (\bibinfo  {publisher} {{Springer Berlin Heidelberg}},\ \bibinfo {address}
  {{Berlin, Heidelberg}},\ \bibinfo {year} {2010})\BibitemShut {NoStop}%
\bibitem [{\citenamefont {Sheka}(2021)}]{sheka2021}%
  \BibitemOpen
  \bibfield  {author} {\bibinfo {author} {\bibfnamefont {D.~D.}\ \bibnamefont
  {Sheka}},\ }\href {\doibase 10.1063/5.0048891} {\bibfield  {journal}
  {\bibinfo  {journal} {Appl. Phys. Lett.}\ }\textbf {\bibinfo {volume}
  {118}},\ \bibinfo {pages} {230502} (\bibinfo {year} {2021})}\BibitemShut
  {NoStop}%
\end{thebibliography}%

\end{document}